\documentclass[11pt,twocolumn]{article}
\usepackage{amsmath} 
\usepackage{authblk}
\usepackage{graphicx,natbib}
\newcommand{\AN}{\textup{\AA}}
\usepackage[hidelinks]{hyperref}

\begin{document}
\onecolumn
\title{CABS-flex predictions of protein flexibility compared with NMR ensembles}
\author[1]{Michal Jamroz}
\author[1]{Andrzej Kolinski}
\author[1]{Sebastian Kmiecik \thanks{sekmi@chem.uw.edu.pl}}

\affil[1]{Faculty of Chemistry, University of Warsaw, Warsaw 02-093, Poland}

\onecolumn
\maketitle
\begin{abstract}
    \textbf{Motivation:} Identification of flexible regions of protein structures is important for understanding of their biological functions.
    Recently, we have developed a fast approach for predicting protein structure fluctuations from a single protein model: the \textsc{cabs}-flex. \textsc{cabs}-flex was shown to be an efficient alternative to conventional all-atom molecular dynamics (\textsc{md}). 
In this work, we evaluate \textsc{cabs}-flex and \textsc{md} predictions by comparison with protein structural variations within \textsc{nmr} ensembles.

\textbf{Results:} Based on a benchmark set of 140 proteins, we show that the relative fluctuations of protein residues obtained from \textsc{cabs}-flex are well correlated to those of \textsc{nmr} ensembles. On average, this correlation is stronger than that between \textsc{md} and \textsc{nmr} ensembles. 
In conclusion, \textsc{cabs}-flex is useful and complementary to \textsc{md} in predicting of protein regions that undergo conformational changes and the extent of such changes.\footnote{\textbf{Availability:} The \textsc{cabs}-flex is freely available to all users at \href{http://biocomp.chem.uw.edu.pl/CABSflex}{http://biocomp.chem.uw.edu.pl/CABSflex}.}
\footnote{\textbf{Contact:} \href{sekmi@chem.uw.edu.pl}{sekmi@chem.uw.edu.pl}}
\footnote{\textbf{Supplementary information:} Supplementary data are available at Bioinformatics online.}

\end{abstract}

\twocolumn
\section{Introduction}

Proteins exist in solution as ensembles of structurally different conformational states. These ensembles can exhibit different degrees of structural diversity, ranging from almost static to highly mobile protein regions. Structural flexibility is one of the key characteristics of proteins, and allows them to play important functional roles in living organisms. Thus, knowledge of conformational states in native-state ensembles can provide important insights into protein functions (e.g. molecular recognition, protein allostery) \citep{Hilser2010b,Fenwick2011a,Gerek2013,Wrabl2011a} as well as protein evolution \citep{Gerek2013,Wrabl2011a}.

Most of the known protein structures have been solved by X-ray crystallography and deposited in the Protein Data Bank (\textsc{pdb}) as a single model. A single crystal structure, however, gives little information about conformational heterogeneity or model accuracy, and this is why the crystallographic community has been urged to deposit an ensemble of solutions whenever possible \citep{Furnham2006a}. An ensemble view of protein structures comes predominantly from \textsc{nmr} spectroscopy which is the method of choice for the determination of protein structure and dynamics in solution \citep{Markwick2008}. \textsc{nmr} spectroscopy routinely provides an ensemble of protein models which usually consists of 20 conformers on average. The precision and accuracy of \textsc{nmr} ensembles have been a subject of a long standing dispute in the field \citep{Spronk2003}. The structure diversity of \textsc{nmr}-derived ensembles may depend not only on the quality and amount of collected data but also on the computational procedures used for generating and selecting low-energy models that fit experimental data. Nevertheless, it has been demonstrated that \textsc{nmr} ensembles may provide valuable insights into protein flexibility that is of practical use in structure-to-function studies \citep{Bolstad2008,Damm2007,Isvoran2011a,Knegtel1997}. Among these studies, particularly interesting is probably the first comparison of \textsc{nmr} ensembles and a collection of crystal structures from the point of using them in structure-based drug design, performed by \citet{Damm2007}. They demonstrated that for human immunodeficiency virus-1 protease (\textsc{hiv}-1p), there is more structural variation between 28 structures in an \textsc{nmr} ensemble than between 90 crystal structures bound to a variety of ligands. Since the \textsc{nmr} ensemble-derived model provided the most general yet accurate representation of the active site of \textsc{hiv}-1p, the authors strongly encourage the use of \textsc{nmr} models in structure-based drug design.

Except for experimental sources, the present views on protein flexibility have been largely obtained thanks to the use of molecular dynamics (\textsc{md}). In the past decades, \textsc{md} has become an indispensable tool for determining conformationally heterogeneous states of proteins, most often through unbiased simulations starting from experimental static structures or in combination with experimental data \citep{Fisette2012a,Vendruscolo2007a}. The idea that unbiased \textsc{md} simulations capture the true dynamic nature of proteins was supported by a study showing that various \textsc{md} force-fields provide a consensus picture of protein fluctuations in solution \citep{Rueda2007}. Using the \textsc{md} simulation data from this study, we very recently demonstrated that the structural and dynamics characteristics of \textsc{md} trajectories are fairly consistent with simulation results from a coarse-grained protein model -- the \textsc{cabs} model \citep{Jamroz2013d}. Importantly, the computational cost of obtaining near-native dynamics by \textsc{cabs} simulations was proved to be much lower (about $6\times 10^3$ times) than that of \textsc{md} (technically, this is the cost of achieving a residue fluctuation profile that best fits that obtained from 10-nanosecond \textsc{md} simulations, see details in \cite{Jamroz2013d}). Following this work, we implemented the developed \textsc{cabs}-model-based protocol for fast simulations of near-native dynamics in a web server called \textsc{cabs}-flex \citep{Jamroz2013e}. 

In previous works, we compared \textsc{cabs}-flex predictions of protein flexibility with a large set of \textsc{md} simulation data \citep{Jamroz2013d,Jamroz2013e}. The comparison tests showed that the \textsc{cabs}-flex method is a computationally efficient alternative to \textsc{md}. The present work describes a comparison of protein fluctuations obtained from \textsc{cabs}-flex and \textsc{md} simulations with fluctuations derived from \textsc{nmr} ensembles.

\section{Methods}

\subsection{Benchmark set}

We used a protein benchmark set constructed and reported by \citet{Jamroz2012a}. The benchmark set contains 140 non-redundant proteins determined by \textsc{nmr} (with \textsc{nmr} ensembles consisting of more than 10 models in their \textsc{pdb} files) and \textsc{md} simulation trajectories deposited in the MoDEL database \citep{Meyer2010a}. The protein set is non-redundant in the sense that it contains no two proteins that have sequence identity higher than a 35\% cutoff according to the \textsc{pisces} database \citep{Wang2003a}.

\subsection{CABS-flex method}

The \textsc{cabs}-flex method follows our earlier work \citep{Jamroz2013d} where we demonstrated that the consensus view of protein near-native dynamics obtained from 10-nanosecond \textsc{md} simulations (all-atom, explicit water, using the four most popular force-fields for all protein metafolds) is consistent with dynamics from the \textsc{cabs} model. The \textsc{cabs}-flex simulation length has been optimized to obtain the best possible convergence with the 10-nanosecond \textsc{md} simulations (see details in \cite{Jamroz2013d}).

\textsc{cabs} is a well-established coarse-grained protein modeling tool for predicting protein dynamics \citep{Kmiecik2012,Kmiecik2007b,Kmiecik2011a} and protein structure \citep{Blaszczyk2013,Kmiecik2007a,Kolinski2005a}. The \textsc{cabs} design is a compromise between high sampling efficiency and high resolution of protein representation. The \textsc{cabs} protein representation is reduced to up to four pseudo-atoms per residue, the force field employs knowledge-based potentials (accounting for solvent effects in an implicit fashion), and the sampling is realized by the Monte Carlo method (details are given in \citet{Kolinski2004a}). The resolution of \textsc{cabs}-generated models allows the reconstruction of physically sound atomistic models \citep{WABICZYSKO,Kmiecik2007a,Kmiecik2012}.

The \textsc{cabs}-based procedure for the simulation of near-native dynamics has been made available as a \textsc{cabs}-flex web server \citep{Jamroz2013e}. The \textsc{cabs}-flex server requires input of a single protein structure and outputs a residue fluctuation profile together with accompanying analysis. Additionally, the \textsc{cabs}-flex pipeline incorporates multiscale reconstruction and optimization procedures \citep{Gront2012a,243} which output an ensemble of protein models (in all-atom resolution) reflecting the flexibility of the input structure.

\subsection{Computing residue fluctuation profiles}

Based on the generated trajectory (\textsc{cabs}-flex or \textsc{md}) or \textsc{nmr} ensemble, superimposed with \textsc{theseus} \citep{Theobald2012}, a residue-fluctuation profile (root mean square fluctuation), is calculated as:

\begin{equation*}
    \mathrm{\textsc{rmsf}} = \sqrt{ \frac{1}{N} \sum_j^N \left( x_i(j) - \left<x_i\right> \right)^2 }
    \label{eq:rmsf}
\end{equation*}

where $<>$ denotes the average over the whole \textsc{nmr} ensemble or trajectory, and $x$ is the position of residue (C$\alpha$ atom) $i$ in the trajectory or \textsc{nmr} ensemble model $j$. 

For the comparison of residue fluctuation profiles obtained from \textsc{cabs}-flex, \textsc{md} and \textsc{nmr} ensembles, we used Spearman's rank correlation coefficient. It quantifies the extent of statistical dependence between pairs of observations (and is better suited to reflect data correlation in the presence of outlier values than the Pearson correlation coefficient). Spearman's rank correlation was also used in our earlier comparisons of \textsc{md} and \textsc{cabs}-flex fluctuation profiles to which we refer in this study \citep{Jamroz2013d,Jamroz2013e}. 

Note that the statistical errors of \textsc{rmsf} values generated by \textsc{cabs}-flex are reflected in root mean squared deviations (\textsc{rmsd}) between \textsc{rmsf} profile data (Figure~\ref{fig:02}B).

\section{Results}

In this work we used a benchmark protein set of 140 proteins collected and reported by \citet{Jamroz2012a}.

\begin{figure}[!tpb]
    \centerline{\includegraphics{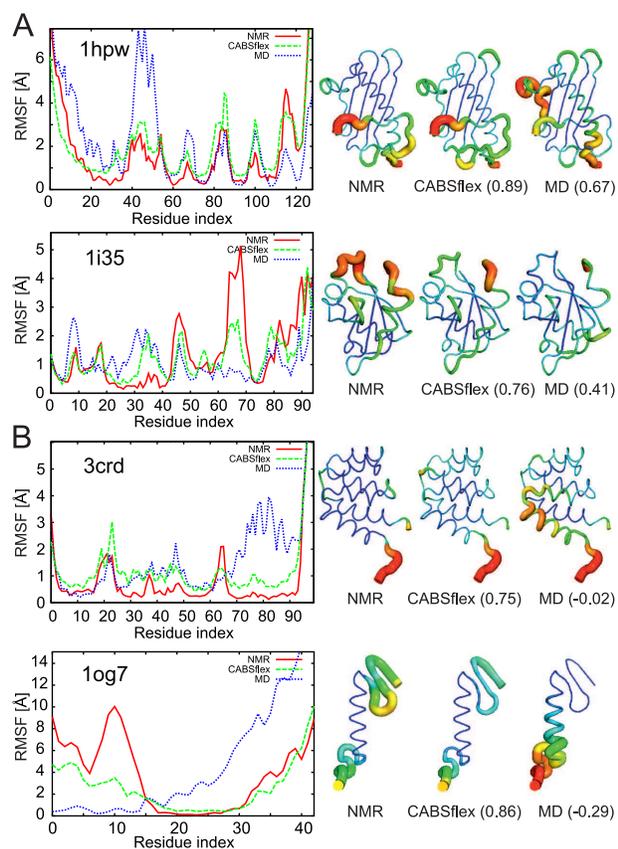}}
\caption{Comparison of residue-fluctuation profiles for example proteins from the benchmark set. The presented examples illustrate several levels of prediction accuracy in comparison to \textsc{nmr} ensembles: (A) high by \textsc{cabs}-flex and average or below average by \textsc{md}, (B) high by \textsc{cabs}-flex and very poor by \textsc{md}. For each protein, residue-fluctuation profiles are visualized on a plot and projected on protein models. The plots present root mean-square fluctuation (\textsc{rmsf}) values (in \AN ngstroms) derived from \textsc{nmr} ensembles (red line) and simulation trajectories: \textsc{cabs} (green line) and \textsc{md} (blue line). The \textsc{rmsf} values are also visualized in the respectively signed protein models (in brackets: correlation coefficients for residue fluctuations between \textsc{nmr} and \textsc{cabs}-flex or \textsc{md}). In the protein models, colors and tube thickness denote \textsc{rmsf} values scaled from the maximum (red color, thick tube) to minimum (blue color, thin tube). Analogous plots for the entire test set are presented in Figure~S1.\label{fig:01}}
\end{figure}

In Figure~\ref{fig:01}, we show a comparison of flexibility for 4 example proteins from the benchmark set. Structural flexibility is presented in the figure as residue-fluctuation profiles, i.e. Root Mean-Square Fluctuation values (\textsc{rmsf}) for each residue (see Methods), visualized in plots or projected on protein models.

\begin{figure}[!tpb]
    \centerline{\includegraphics{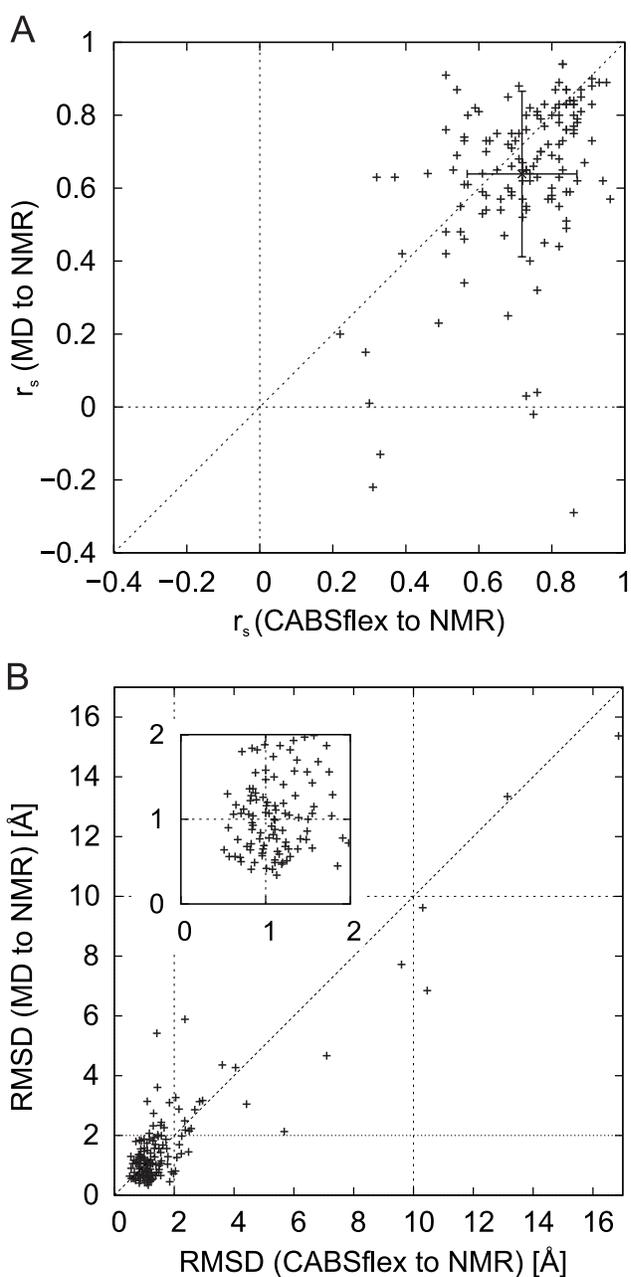}}
    \caption{Comparison of residue-fluctuation profiles for the benchmark set. For the set of 140 protein structures, a comparison between \textsc{cabs}-flex and \textsc{nmr} is presented together with that of \textsc{md} and \textsc{nmr}. For each protein, residue-fluctuation profiles (root mean squared fluctuations, \textsc{rmsf}) are compared using: (A) Spearman's correlation coefficient ($r_s$), and (B) average \textsc{rmsd} (root-mean square deviation) values.\label{fig:02}}
\end{figure}

In Figure~\ref{fig:02}, we present a comparison of residue-fluctuation profiles for the entire benchmark set. The comparison is done using Spearman's correlation coefficient ($r_s$) (Figure~\ref{fig:02}A) and average root mean square deviation (\textsc{rmsd}) between \textsc{rmsf} values of \textsc{md}/\textsc{nmr}/\textsc{cabs}-flex (Figure~\ref{fig:02}B). Remarkably, the average $r_s$ between \textsc{cabs}-flex and \textsc{nmr} ensembles is slightly less scattered than that between \textsc{md} and \textsc{nmr}: $0.72 (\pm 0.15)$ and $0.64 (\pm 0.23)$, respectively (standard deviation values are given in brackets). 

The $r_s$ correlation coefficient is a measure of statistical dependence between compared residue-fluctuation profiles and does not reflect differences in profile amplitudes. This is reflected in the average \textsc{rmsd} between the compared profiles shown in Figure~\ref{fig:02}B. As presented in the plot, the \textsc{rmsd} between \textsc{nmr} profiles and \textsc{cabs}-flex or \textsc{md} profiles usually does not exceed 2 \AN. In general, the higher structural heterogeneity in \textsc{nmr} ensembles, the higher the presented \textsc{rmsd} values. The largest \textsc{rmsd} values correspond to proteins with highly flexible regions. For instance, the highest \textsc{rmsd} values (\textsc{nmr} to \textsc{md} as well as \textsc{cabs}-flex to \textsc{nmr}) correspond to the structure of \textsc{cide}-N Domain of \textsc{cide}-B protein (\textsc{pdb} ID: 1d4b) which has largely disordered regions of substantial length (residues 1-31 and 111-122).
The exact $r_s$ and \textsc{rmsd} values for each protein are given in Table S1 together with accompanying data.

\section{Discussion}

The proteins from the benchmark set represent different degrees of structural variability within \textsc{nmr} ensembles. The degree of variability (average displacement per residue) ranges from 0.2 to almost 12 \AN. For the entire benchmark set, the average displacement per residue in \textsc{nmr} ensembles is 1.68 \AN\ (the values for each protein are given in Table S1). 

The analysis of variability of \textsc{nmr} ensembles vs. prediction quality showed a tendency that the higher the flexibility observed in an \textsc{nmr} ensemble, the better the correlation coefficient ($r_s$) between \textsc{nmr} and \textsc{cabs}-flex or \textsc{md} fluctuation profiles. For 57\% of proteins from the benchmark set, the average displacement within their \textsc{nmr} ensembles is higher than 1 \AN. In this subset, the average $r_s$ between \textsc{nmr} and simulation (\textsc{cabs}-flex or \textsc{md}) is slightly higher (0.78 for \textsc{cabs}-flex and 0.69 for \textsc{md}) than for proteins with less variable \textsc{nmr} ensembles (see Table \ref{tab:01}). 
\begin{table*}[htb]
    \caption{Average Spearman's correlation coefficients ($r_s$) between residue-fluctuation profiles. \label{tab:01}}
    {\begin{tabular}{lp{0.7in}p{0.7in}p{0.7in}}
        &    \multicolumn{3}{c}{Benchmark dataset of \textsc{nmr}-solved proteins}\\\cline{2-4}

& Entire dataset (140 proteins) & Subset with \textsc{rmsd} of \textsc{nmr} ensemble $\le$1~\AN\  (60 proteins) & Subset with \textsc{rmsd} of \textsc{nmr} ensemble $>$1~\AN\  (80 proteins)\\
\textsc{cabs}-flex vs. \textsc{nmr} & $\mathbf{0.72 (\pm 0.15)}$ & $0.64 (\pm0.17)$ & $0.78 (\pm 0.11)$ \\
\textsc{md} vs. \textsc{nmr}        & $\mathbf{0.64 (\pm 0.23)}$ & $0.57 (\pm 0.25)$ &  $0.69 (\pm 0.19)$\\
\textsc{cabs}-flex vs. \textsc{md}  & $\mathbf{0.67 (\pm 0.18)}$ & $0.64 (\pm 0.17)$ & $0.69 (\pm 0.17)$\\ 
\end{tabular}}

{The table shows an average pairwise comparison between \textsc{cabs}-flex, \textsc{md} and \textsc{nmr} ensembles. The average correlation values (and standard deviations in brackets) are presented for the entire protein benchmark set and its subsets having average fluctuations in the NMR ensemble: lower (\textsc{rmsd} $\le$ 1~\AN) or higher (\textsc{rmsd} $>$1~\AN).}
\end{table*}

Furthermore, we examined another subset of proteins for which \textsc{cabs}-flex predictions were the poorest (with $r_s < 0.5$: 1k8b, 1waz, 1kkg, 1k5k, 1cok, 1sgg, 1pcp, 1pav, 1p6q, 2rgf). In this subset of 10 proteins, the average $r_s$ between \textsc{nmr} and \textsc{cabs}-flex fluctuation profiles was 0.35, while that between \textsc{nmr} and \textsc{md} was even lower: 0.26. The subset analysis showed that 9 out of 10 proteins had \textsc{nmr} ensembles exhibiting almost no or small flexibility, in contrast to \textsc{cabs}-flex or \textsc{md} predictions (the exception was 1pcp which has a small amount of secondary structure only). For these 9 proteins, the average displacement per residue within \textsc{nmr} ensembles was below 0.5 \AN\ (counted for the entire or most of the chain). Such large rigidity does not seem to be justified by the structural characteristics of these proteins. For at least some of them, highly homologous counterparts can be found in the Protein Data Bank which show more structural variation than the analyzed \textsc{nmr} ensembles.

The above observations suggest that an important source of poor correspondence between fluctuations from computational predictions (from \textsc{cabs}-flex or \textsc{md}) and \textsc{nmr} ensembles is the underestimation of fluctuations in \textsc{nmr} ensembles. Several studies strongly indicate that fluctuations in \textsc{nmr} ensembles are underestimated and do not reflect real structural heterogeneity \citep{Pfeiffer1997,Scheek1995,Spronk2003,Torda1990}. The underestimations are largely due to shortcomings of computational procedures used to generate the ensembles based on \textsc{nmr} data.

The \textsc{cabs}-flex method provides an   alternative to other efficient computational tools generating protein residue fluctuation profiles, such as sequence-based predictors of protein disordered regions \citep{MESZAROS_new} or coarse-grained normal mode analysis \citep{MA_new}. 
Most disorder prediction algorithms (such as \textsc{disopred}, \cite{WARD_new}) perform well for stable, globular domains, or highly flexible disordered regions without a strong structural preference. 
However, their performance does not meet expectations for structurally ambiguous regions \citep{MESZAROS_new}. 
Therefore, in comparison to sequence-based disorder prediction algorithms, \textsc{cabs}-flex is better suited to detecting non-obvious dynamic behavior (e.g. significant fluctuations within the well-defined secondary structural elements that could be of biological importance). 
Another class of commonly used algorithms that compute protein fluctuation profiles use normal mode analysis (\textsc{nma}) based on elastic network models or other coarse-grained models (e.g. WEBnma server, \cite{HOLLUP_new}). 
In comparison to elastic network models, \textsc{cabs}-flex uses more detailed information on the protein system and generates residue fluctuation profiles better correlated (on average) with those obtained by all-atom \textsc{md} (see our discussion in \cite{Jamroz2013d}). The \textsc{cabs}-flex generated models (or trajectory) can also be subjected to \textsc{nma}. 
As we demonstrated earlier \citep{Jamroz2013d}, essential movements derived from \textsc{cabs}-flex trajectories might not be accurate individually, but when considered together they provide a similar description to that obtained by all-atom \textsc{md}. Readers interested in applying the \textsc{nma} may refer to a review on the usefulness and limitations of the method \citep{MA_new}.

\section{Conclusion}

Due to the dynamic nature of proteins, structure-based studies of protein functions require accurate description of protein flexibility.

Crystallographic B-factors are perhaps the most common measure used for the elucidation of residue fluctuations, and this is probably because the majority of known structures have been solved by X-ray crystallography. The B-factors reflect protein flexibility but are also influenced by crystallization conditions, the refinement method (used for the interpretation of X-ray data) and, importantly, the molecular environment of the crystal structure. The crystal environment has a significant effect on protein flexibility: the spectrum of fluctuations is considerably flattened in crystal as compared with that in solution \citep{Eastman1999a}. Moreover, most of X-ray structures have been determined at cryogenic temperatures. Crystal cryo-cooling has been shown to reduce B-factors, introduce packing defects, and it may result in unrealistically unique nonfunctional structures \citep{Fraser2011,Rasmussen1992}. Therefore, descriptions of protein flexibility derived from X-ray models and B-factors must be approached with caution.

\textsc{nmr} and all-atom \textsc{md} are now the methods of choice for investigation of protein flexibility in solution. 
Because of the difficulty of \textsc{nmr} studies and timescale problems in all-atom \textsc{md}, coarse-grained methods have emerged as an inexpensive and powerful alternative. 
The design of coarse-grained methods successfully applied for large time-scale investigations of protein dynamics encompasses entirely different modeling strategies \citep{MAISURADZE_new,EMPERADOR_new,Jamroz2013d}. An excellent review on the successes and shortcomings of diverse coarse-grained representations of protein flexibility is provided in \citep{OROZCO_new}. 

In this work, we compare \textsc{cabs}-flex predictions of protein fluctuations with that of derived from \textsc{nmr} ensembles and \textsc{md} simulations. The comparison shows that \textsc{cabs}-flex produces, on average, a more similar distribution of residue fluctuations to \textsc{nmr} ensembles than \textsc{md} does.
This is due to more efficient sampling compared to \textsc{md}, which leads to additional fluctuations or fluctuation amplitudes that better fit the \textsc{nmr} ensemble data. 
Moreover, the results from \textsc{cabs}-flex and \textsc{md} can complement each other in the sense that the flexibility of some protein regions may be better retrieved by one of these methods, while the remaining part by the other one. In summary, our results suggest that for the accurate assessment of protein flexibility it is reasonable to analyze results from both \textsc{cabs}-flex and atomic \textsc{md} simulations. Since the \textsc{cabs}-flex method provides a significantly cheaper means of accessing backbone dynamics than atomic \textsc{md}, it is a promising tool for larger and/or initial reconnaissance screening studies, for example of the effect of mutations on protein stability or structure-based drug design. 

\section*{Acknowledgement}

\paragraph{Funding} 
The authors acknowledge funding from a Foundation for Polish Science TEAM project [TEAM/2011-7/6]
co-financed by the EU European Regional Development Fund operated within the Innovative Economy
Operational Program; Polish National Science Centre [NN301071140]; Polish Ministry of Science and
Higher Education [IP2011 024371].

\bibliographystyle{apalike}
\bibliography{biblioteka}

\end{document}